\documentclass[conference]{IEEEtran}
\usepackage{packages}

\IEEEoverridecommandlockouts
\IEEEpubid{\makebox[\textwidth]{ISBN 978-3-903176-63-8\copyright{} 2024 IFIP \hfill}}%

\iftrue
\makeatletter
\let\old@ps@headings\ps@headings
\let\old@ps@IEEEtitlepagestyle\ps@IEEEtitlepagestyle
\def\confheader#1{%
    \def\ps@IEEEtitlepagestyle{%
        \old@ps@IEEEtitlepagestyle%
        \def\@oddhead{\strut\hfill#1\hfill\strut}%
        \def\@evenhead{\strut\hfill#1\hfill\strut}%
    }%
    \ps@headings%
}
\makeatother

\confheader{%
    \color{TUMBlue}
    \fbox{
        \begin{minipage}{2\linewidth}
            \footnotesize
        If you cite this paper, please use the IFIP Networking reference: M. Kempf, N. Gauder, B. Jaeger, J. Zirngibl, G. Carle. 2024. A Quantum of QUIC: Dissecting Cryptography with Post-Quantum Insights. \textit{In Proc. of IFIP Networking Conference (IFIP Networking)}. IFIP.
        \end{minipage}
}
}
\fi

\begin{document}

\title{A Quantum of QUIC: Dissecting Cryptography with Post-Quantum Insights}

\author{\IEEEauthorblockN{Marcel Kempf, Nikolas Gauder, Benedikt Jaeger, Johannes Zirngibl, Georg Carle}
\IEEEauthorblockA{\textit{Technical University of Munich, Germany}\\
\{kempfm, gauder, jaeger, zirngibl, carle\}@net.in.tum.de\\
}
}

\maketitle

\begin{abstract}
\quic is a new network protocol standardized in 2021.
It was designed to replace the \tcptls stack and is based on \ac{udp}.
The most current web standard \hthree is specifically designed to use \quic as transport protocol.
\quic claims to provide secure and fast transport with low-latency connection establishment, flow and congestion control, reliable delivery, and stream multiplexing.
To achieve the security goals, \quic enforces the usage of \tlsonethree.
It uses \ac{aead} algorithms to not only protect the payload but also parts of the header.
The handshake relies on asymmetric cryptography, which will be broken with the introduction of powerful quantum computers, making the use of post-quantum cryptography inevitable.

This paper presents a detailed evaluation of the impact of cryptography on \quic performance.
The high-performance \quic implementations \lsquic, \quiche, and \msquic are evaluated under different aspects.
We break symmetric cryptography down to the different security features.
To be able to isolate the impact of cryptography, we implemented a \noop \ac{aead} algorithm which leaves plaintext unaltered.
We show that \quic performance increases by 10 to \sperc{20} when removing packet protection.
The header protection has negligible impact on performance, especially for \acs{aes} ciphers.
We integrate post-quantum cryptographic algorithms into \quic, demonstrating its feasibility without major changes to the \quic libraries by using a \ac{tls} library that implements post-quantum algorithms.
\kyber, \dilithium, and \falcon are promising candidates for post-quantum secure \quic, as they have a low impact on the handshake duration.
Algorithms like \sphincs with larger key sizes or more complex calculations significantly impact the handshake duration and cause additional issues in our measurements.

\end{abstract}

\acresetall


\section{Introduction}
\label{sec:introduction}

\quic is a new transport protocol designed to improve on the widely used \tcptls stack, standardized by the \ac{ietf} in 2021~\cite{rfc9000}.
It is the basis for new protocols like \hthree and \ac{masque}, which powers Apple's private relay service.
Like \ac{tcp}, it is connection-oriented, reliable, and features flow and congestion control.
Additionally, \quic has numerous advantages over \ac{tcp}, \eg{} support for connection migration, stream multiplexing, and always-on encryption.
To achieve the latter, \ac{tls}~1.3 is strictly integrated into \quic{}~\cite{rfc9001}.
The \quic{} handshake combines both the transport and \ac{tls} handshake, which allows fast connection establishment.
Furthermore, it encrypts all following payload and adds additional header protection.

These properties are desirable in many use cases.
However, always requiring \ac{tls} is often criticized for inducing additional overhead in scenarios where those properties are not required~\cite{quic-crypto-discussion, quic-crypto-discussion-2}.
\citet{jaeger2023quic} have shown that crypto is the second most expensive component of \quic{} besides packet I/O.

Besides the effect of symmetric cryptography on performance during the connection, the integration and performance evaluation of quantum-safe algorithms in QUIC has not been evaluated in detail.
Traditional asymmetric cryptography, which is used during the \quic handshake, will be broken with the introduction of powerful quantum computers.
The \ac{nist} has been working on selecting quantum-safe cryptographic algorithms for standardization since 2017~\cite{pqc-standardization}.

In this work, we perform a detailed evaluation of the impact of cryptography on \quic{} performance.
We analyze the impact of symmetric cryptography in form of packet and header protection during a connection.
Different post-quantum key exchange and signature algorithms are integrated into two \quic{} implementations to evaluate the performance impact on the handshake.

Our key contributions in this work are:

\first{} We evaluate the impact of cryptography on \quic{} performance for different libraries in detail.
We differentiate between the impact of payload encryption and header protection in a controlled environment.

\second{} We analyze if larger \acp{mtu} can mitigate the impact of encryption and improve the performance of \quic{}.
This is especially relevant in controlled environments, \eg{} in-datacenter networks.

\third{} We integrate quantum-safe cryptographic algorithms into two of the chosen \quic{} implementations to evaluate the performance impact.
We dissect the handshake to precisely locate performance bottlenecks and limitations.

\fourth{} We publish versions of BoringSSL and OpenSSL with \noop ciphers following the required interface.
These can be used to evaluate the impact of cryptography within other \quic{} libraries or to remove the impact of cryptography for other evaluations.

\vspace{0.2em}
We explain relevant background regarding cryptography in \quic in \Cref{sec:background} and cover related work in \Cref{sec:related-work}.
In \Cref{sec:approach}, we introduce our approach and the measurement setup.
Our evaluations are presented in \Cref{sec:evaluation}.
Finally, the main findings are concluded in \Cref{sec:conclusion}.

\section{Background}
\label{sec:background}

This section introduces relevant background about cryptography in QUIC, followed by an overview of \ac{pqc} and how \quic is affected when integrating \ac{pqc}.

\subsection{QUIC Cryptography}

\begin{figure}[t]
    \centering
    \includegraphics[width=\linewidth]{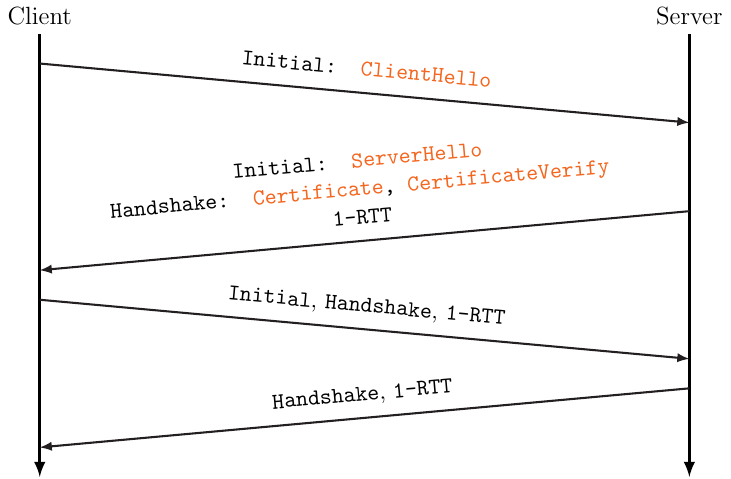}
    \caption{Simplified illustration of a \quic 1-RTT handshake.}
    \label{fig:quic-handshake}
\end{figure}

\quic includes always-on encryption with \tlsonethree.
While asymmetric cryptography is used during the handshake, symmetric cryptography is used during the connection.
In \Cref{fig:quic-handshake}, a simplified illustration of a \quic handshake is shown.
Asymmetric cryptography is only happening in the orange parts.
It is important to note that all shown handshake components may be spread over multiple \quic packets.
This happens especially with \ac{pqc}, where the certificate is too large to fit into a single packet.

For performance and security considerations, \tlsonethree limits the amount of available ciphers to only \ac{aead} algorithms~\cite{rfc8446}.
They are designed to encrypt data while applying integrity protection to the data itself and also additional metadata in one single pass.
During the connection, the packet payload is encrypted, and the header is integrity-protected along with the payload.
All non-\ac{aead} algorithms have been pruned from the standard and only five are available for \tlsonethree.
\quic further limits this set to four:
\texttt{AES\_128\_GCM},
\texttt{AES\_128\_CCM},
\texttt{AES\_256\_GCM},
and \texttt{CHACHA20\_POLY1305}~\cite{rfc9001}.
The \ac{aes} algorithms are block ciphers which profit from hardware acceleration on modern \texttt{x86} \acp{cpu}~\cite{intelaesni}, while \ac{chacha20}~\cite{rfc7539} is a stream cipher performing well when hardware acceleration is lacking.

\begin{figure}[t]
    \centering
    \includegraphics[page=4, width=\linewidth]{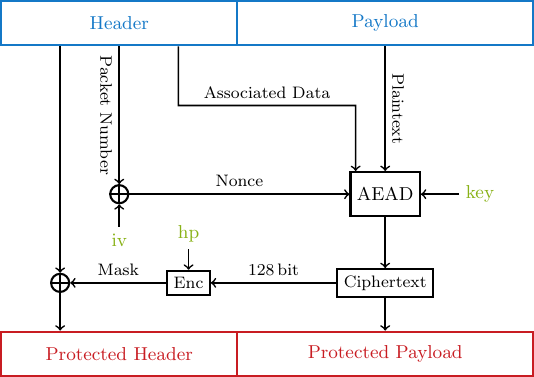}
    \caption{Packet and header protection in \quic using the initialization vector (iv), the header protection key (hp), and the \quic key (key). All three are derived from the connection's TLS secrets.}%
    \label{fig:quic-encryption}
\end{figure}

Besides protecting the payload, parts of the header are also encrypted.
To prevent network ossification and ensure header authenticity, all fields not required for decryption are protected during the connection.
This includes the packet number and several bits in the header.
\Cref{fig:quic-encryption} shows the way a \quic packet is protected and which keys are involved.
First, the packet protection is applied by encrypting the payload, using the \ac{aead} cipher.
The header serves as additional data and is not encrypted but authenticated.
The nonce is derived by XORing the packet number with the initialization vector, ensuring that the nonce is unique for every packet.
From the resulting ciphertext of the \ac{aead} encryption, a \qty{128}{\bit} sample is taken and used as input to a cipher.
The \ac{aead} algorithm for the packet protection determines the respective cipher.
If an \ac{aes} cipher suite is used, the respective \ac{aes} cipher is applied in \ac{ecb} mode.
For the ChaCha20 cipher suite, the raw ChaCha20 function is used.
From the ciphertext of this single encryption call, \qty{8}{\byte} are used to mask the fields to be protected in the header~\cite{rfc9001}.

\subsection{Quantum-Safe Cryptography}

Traditional asymmetric cryptography in \ac{tls} is based on prime factorization or the (elliptic-curve) discrete logarithm problem.
Both can be solved with powerful quantum computers using Shor's algorithm~\cite{shor1999polynomial}.
Using Grover's algorithm~\cite{10.1145/237814.237866}, symmetric keys can be brute-forced more efficiently, halving the security level in \unit{\bit}.
However, symmetric cryptography is not as vulnerable, as larger keys can mitigate this problem.
Hence, \ac{pqc} introducing new quantum-secure cryptographic systems is needed as a replacement.
In light of this, the \ac{nist} launched the \textit{Post-Quantum Cryptography Standardization} program in 2017 to standardize quantum-secure cryptographic primitives~\cite{pqc-standardization}.

In this work, we focus on the following post-quantum key exchange and signature algorithms:
Kyber, BIKE, HQC, Dilithium, Falcon, and \sphincs.
Detailed information about these algorithms can be found in the \textsl{liboqs} documentation from \ac{oqs}~\cite{oqs-algorithms}.
For the ones selected for standardization, the \ac{nist} has assigned new names, \eg{} ML-KEM for \kyber~\cite{cf-pq-blog-2024}.
In this paper, we use the old names for the algorithms, as they are more commonly known.
The algorithms are grouped into hash-, code-, and lattice-based algorithms, each coming with unique advantages and limitations~\cite{10.5555/1522375}.

\ac{nist} also established different quantum security strength categories to compare various algorithms regarding their security.
Relevant to our work are the so-called \ac{nist} levels I, III, and V.
Algorithms in \ac{nist} level I are at least as hard to break as AES-128 through exhaustive key search.
The \ac{nist} levels III and V correspond to the strength of AES-192, respective AES-256.

\section{Related Work}
\label{sec:related-work}

\citet{jaeger2023quic} performed a broad and comprehensive performance evaluation of \quic{} libraries.
Among other implementations, they evaluated \lsquic~\cite{lsquic} and \quiche~\cite{quiche} as well as \ac{tcp} over \ac{tls}.
In their comparison of the goodput they revealed that the performance of hardware accelerated \ac{aes} is superior to \ac{chacha20}.
In their \ac{cpu} utilization measurements, they found that cryptographic operations contribute between \sperc{10} and \sperc{20} to the total \ac{cpu} utilization.
We further extend the evaluation with \msquic~\cite{msquic} and focus on the goodput for the different cipher suites rather than on the effect of hardware acceleration.
We also provide a fine-grained analysis of the different components of the symmetric cryptography, \ie{} the header and packet protection, and the impact of the \ac{mtu} on the performance of \quic and cryptography.
Lastly, we integrate quantum-safe cryptographic algorithms to evaluate the performance impact on the handshake and identify possible problems when integrating \ac{pqc} into \quic.

\citet{yang2020quicnicoffloading} analyzed different \quic{} implementations in the context of \ac{nic} offloading, aiming to define a set of primitives that a \ac{nic} should offer to efficiently offload \quic.
They looked at the following implementations of \quic: \quant~\cite{quant}, \quicly~\cite{quicly}, \picoquic~\cite{picoquic}, and \mvfst~\cite{mvfst}.
Like \citeauthor{jaeger2023quic}, they showed that high costs are associated with crypto: up to \sperc{40} of the \ac{cpu} usage is attributed to cryptographic operations (\quant).
More specifically, they discovered that around \qtyrange[range-phrase = --]{75}{80}{\percent} of the crypto-related \ac{cpu} usage is associated with \ac{aead} function calls.
We analyze the cryptographic operations more thoroughly and break them down in more detail to show the impact of \quic's security features, \ie{} the header and packet protection.

Apart from the works previously presented, various papers deal with the analysis of \quic itself without focusing specifically on cryptography \cite{bauer2023tcpandquic, yu2021quicproductionperformance}. %
They performed comparative studies on the performance of \ac{tcp} and \quic, \eg{} \citet{yu2021quicproductionperformance} ran their tests under different network conditions and workloads against production endpoints from \google, \cloudflare, and \facebook.

\citet{marx2020implementationdiversity} researched \quic{} features in 15 \ac{http}/3 implementations, \eg{} flow and congestion control, stream multiplexing, and the 0-\ac{rtt} handshake.
They summarize that there are significant differences regarding the quality of the \quic{} implementations and that most of them are not completely optimized, wasting potential performance gains.
However, in this work, the focus is on the cost of cryptography.

While some studies evaluated path \ac{mtu} discovery in general or whether it is implemented (\eg{} \citet{marx2020implementationdiversity}), to the best of our knowledge, no study evaluated the impact of larger \acp{mtu} on \quic{} performance and its relation to cryptography.

\citet{sosnowski2023performance} investigated the performance implications of using post-quantum algorithms in \tlsonethree handshakes over \ac{tcp}.
Their results reveal that PQ algorithms (including hybrids) can be faster than the state-of-the-art in ideal network conditions.
However, in low-bandwidth environments, the increased data usage becomes a bottleneck.
Moreover, they found that the large key sizes can cause unwanted side effects, e.g., additional \acp{rtt} due to the slow start phase of the \ac{tcp} congestion control.
We use \quic instead of \ac{tcp} and also evaluate the integration into \quic libraries and the corresponding issues.

\citet{10.1007/978-3-031-22390-7_6} analyzed the impact of \ac{pqc} on the \quic handshake.
They found that the handshake takes longer with increasing security levels or worse network conditions.
They only looked at the two lattice-based signature algorithms \dilithium and \falcon and did not analyze key encapsulation mechanisms, what we do in this work.

\section{Approach}
\label{sec:approach}

In this section, we present our approach to conduct measurements and evaluate collected metrics.
After introducing the selected \quic libraries, we present the adjustments made to the \ac{tls} libraries to allow for measurements without cryptography.

\subsection{Measurement Framework}

To execute our measurements in a reproducible manner, we extended the adapted \quic Interop Runner presented by \citet{jaeger2023quic}.
This framework was built to orchestrate measurements on bare-metal servers and to provide a reproducible environment for \quic measurements with extensible configuration and logging capabilities.
It is based on the \quic Interop Runner presented by \citet{seemann2020automating}, which was initially designed to perform interoperability tests between different \quic implementations.
We added features to change the path \ac{mtu} and modified the build process to include our custom \ac{tls} libraries presented in \Cref{sec:approach-ssl}.

\subsection{Hardware Configuration}

All machines used for the measurements are equipped with an AMD EPYC 7543 32-Core \ac{cpu}, \SI{512}{\giga\byte} memory, and a 10GBASE-T Broadcom BCM57416 \ac{nic}.
We use Debian Bullseye on 5.10.0-8-amd64 as the operating system without additional configurations.

\subsection{Implementations}

For our evaluation, we consider \lsquic~\cite{lsquic}, \quiche~\cite{quiche}, and \msquic~\cite{msquic}, as those implementations are widely used for production \quic servers~\cite{zirngibl2023quic}.
Related work also showed that these implementations perform better than the majority of other tested \quic implementations~\cite{jaeger2023quic, koenig2023quic}. %
We use the example server and client applications provided by the respective libraries for interop testing.
While we configured \lsquic and \quiche to use \ac{http}/3, \msquic provides only an \ac{http}/0.9 implementation.
For the \msquic measurements, we used the \texttt{QUIC\_PARAM\_CONN\_DISABLE\_1RTT\_ENCRYPTION} connection parameter to disable the 1-\ac{rtt} encryption completely.
We refer to this configuration as \texttt{NOENC} in the following.
As this feature is only for testing and performance evaluation, a constant must be defined to enable the so called "insecure features".
Additionally, we use a \tcptls stack consisting of a server using \textsl{nginx} and a client using \textsl{curl} for comparison.

\subsection{Adjustments to \boringssl and \openssl}
\label{sec:approach-ssl}

\quic is strongly coupled with \ac{tls} encryption and is generally not designed to operate without it.
Completely removing cryptography from a \quic implementation requires major adjustments to the library.
Therefore, we opted for implementing a \noop cipher for \boringssl~\cite{boringssl} and \openssl~\cite{openssl} which just returns the plain text as cipher text and thus does not perform any cryptographic operations.
This approach allows us to easily perform measurements with other \quic libraries using \boringssl or \openssl as the \ac{tls} library.

Our cipher suite \texttt{TLS\_NOOP\_SHA256} uses SHA-256 for hashing and the custom \texttt{NOOP} algorithm for encryption and decryption.
The asymmetric part during the handshake remains unchanged.
As the algorithm for header protection depends on the selected \ac{aead} algorithm~\cite[Section 5.4.1]{rfc9001}, we decoupled it from the \ac{aead} algorithm in the used \quic libraries and thus are able to choose the header protection algorithm.
This approach allows for easy integration and evaluation of other \quic libraries with our custom \ac{tls} libraries.
The patched \boringssl and \openssl libraries are available on GitHub~\cite{tls-lib-publication}.
To be able to also perform measurements with quantum-resistant cryptographic algorithms, we also integrated our changes into the \boringssl fork of the \ac{oqs} project~\cite{oqs-boringssl}.
The quantum-safe key exchange and signature algorithms are included via the \textsl{liboqs} library, which also originates from the \ac{oqs} project.

\section{Evaluation}%
\label{sec:evaluation}

\begin{figure*}
    \centering
    \includegraphics[width=\linewidth]{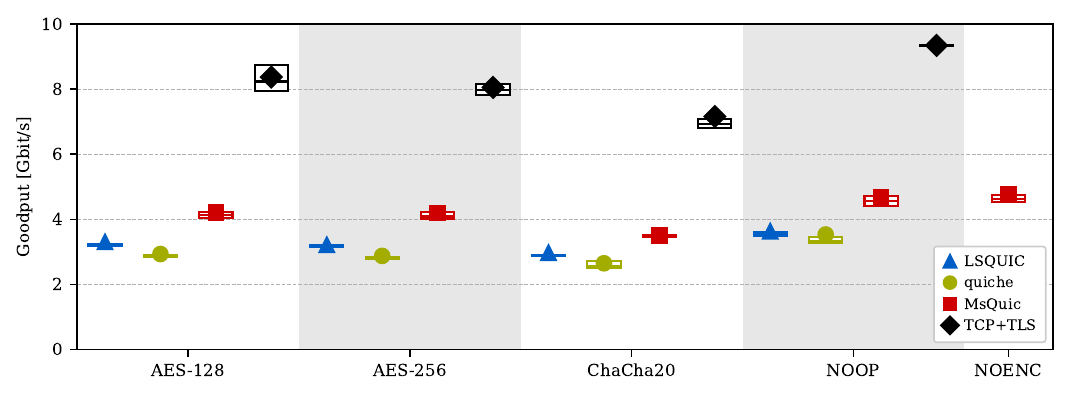}
    \caption{Goodput achieved with different AEAD ciphers.}%
    \label{fig:ciphers}
\end{figure*}

With the previously presented measurement framework and the modified \quic implementations and \ac{tls} libraries, we evaluate the cost of cryptography in \quic.

Besides evaluating the different \ac{aead} algorithms and comparing them with our \noop implementation, we also evaluate the impact of \quic's header protection on the goodput.
The cost of the different cryptographic operations is evaluated in detail with profiling output.

After analyzing symmetric cryptography, we benchmark post-quantum key exchange and signature algorithms integrated into \lsquic and \quiche.
Also looking at hybrid approaches, the impact on the handshake latency is evaluated.

In every measurement, the client downloads an \SI{8}{\gibi\byte} file over \ac{http}.
To ensure a large enough sample, every measurement was repeated 25 times and the average was taken.
We also set the \ac{udp} receive buffer size to \SI{6656}{\kibi\byte} which is 32 times the default size of \SI{208}{\kibi\byte}.
\citet{jaeger2023quic} have shown that the default buffer size is too small for high-rate links.
The congestion control algorithm was fixed to \texttt{cubic} for all measurements.
If not stated otherwise, no header protection is applied in the measurements with the \noopcipher.
All shown boxplots use a horizontal line for the median and an icon such as $\blacktriangle$ for the mean.
The boxes are drawn from quartile $Q_1$ to $Q_3$.

\subsection{AEAD Algorithms}

Comparing \ac{tls} over \ac{tcp} with \quic highlights how \quic's use of \ac{udp} affects performance.
\Cref{fig:ciphers} shows the goodput achieved with different \ac{aead} ciphers \texttt{AEAD\_AES\_128\_GCM}, \texttt{AEAD\_AES\_256\_GCM}, and \texttt{AEAD\_CHACHA20\_POLY1305} as well as our \noop implementation.
\ac{tls} over \ac{tcp} consistently outperforms all tested \quic implementations across all \ac{aead} algorithms, likely due to \ac{tcp}'s mature optimizations.

There are no noticeable differences in performance between the 128-bit and the 256-bit \ac{aes} cipher.
The \ac{aead} algorithm \texttt{AEAD\_CHACHA20\_POLY1305} using the ChaCha20 stream cipher is about \SIrange{9}{16}{\percent} slower than \ac{aes}-based algorithms in combination with hardware acceleration.
As it was shown by \citeauthor{jaeger2023quic}, ChaCha20 shows a significant performance improvement over \ac{aes} when hardware acceleration is not available and is therefore a valuable alternative for endpoints with hardware constraints~\cite{jaeger2023quic}.
With the \noopcipher, the achieved goodput increases between \sperc{10} and \sperc{20} for the tested \quic implementations and \sperc{12} for \ac{tls} over \ac{tcp}.

We also measured the goodput of the \msquic implementation with \noenc.
This configuration completely disables the 1-\ac{rtt} encryption and thus the cryptographic operations.
In comparison to the \noopcipher, the goodput increases by \sperc{13.6} to \SI{4782}{\Mbps}.
The main difference between the two is that \noenc does not even call \openssl for the protection of 1-\ac{rtt} packets, while \noop still calls \openssl performing operations like \texttt{memcpy()}.
This performance impact is also analyzed in \Cref{sec:evaluation-perf}, where we take a closer look at the \ac{cpu} utilization of the different cryptographic operations.

\subsection{Header Protection}
\label{sec:hp-perf}

With a combination of the options to select the \ac{aead} algorithm and the header protection algorithm, we can evaluate the impact of header protection on the goodput.

\begin{figure}
    \centering
    \includegraphics[width=\linewidth]{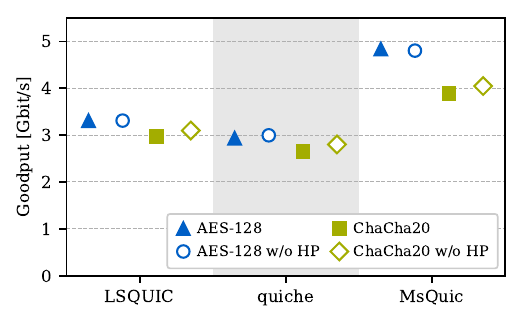}
    \caption{Impact of header protection for different \ac{aead} algorithms and \quic implementations on the goodput.}
    \label{fig:hp}
\end{figure}

\Cref{fig:hp} shows the goodput achieved with different \ac{aead} algorithms, with and without header protection.
As it can be seen, the \ac{aes} header protection has a negligible impact on the goodput.
While it has less than \sperc{1} impact for \lsquic and \msquic, the goodput of \quiche is slightly increased by \sperc{2}.
The header protection with ChaCha20 has a higher impact on the goodput, increasing it by \SIrange{4}{6}{\percent}.
As ChaCha20 showed slightly lower goodput than \ac{aes} in \Cref{fig:ciphers} already, this observation is not surprising.
It can be concluded that the header protection, especially for \ac{aes}, is virtually free and does not have a significant impact on the goodput.

\subsection{CPU Time Consumption for Packet and Header Protection}
\label{sec:evaluation-perf}

To understand what limits the goodput and where bottlenecks are, we take a closer look at the \ac{cpu} utilization of client and server.
We use \texttt{perf} in combination with a custom mapping to retrieve the number of perf samples for packet and header protection.

\Cref{tab:perf} shows the distribution of samples for the packet and header protection mechanisms in relation to the total number of samples belonging to the respective \quic implementation.
Each implementation was tested with \texttt{AEAD\_AES\_128\_GCM}, \texttt{AEAD\_CHACHA20\_POLY1305}, and our \noopcipher for a better comparison.
AES and ChaCha20 use their respective header protection algorithm.
For \msquic, we also included results from measurements with \noenc.
As \ac{aes}-256 again shows similar performance to \ac{aes}-128, it is not included in the table.

\begin{table}[bt]
	\centering
    \caption{Distribution of \texttt{perf} samples for the packet protection (PP) and header protection (HP) mechanisms in relation to the total number of samples belonging to the respective \quic implementation.}
	\label{tab:perf}
    \begin{tabular}{l l r r r r}
		\toprule
         & & \multicolumn{2}{c}{\textbf{Client}} & \multicolumn{2}{c}{\textbf{Server}} \\
        \textbf{Cipher} & \textbf{Impl.} & \textbf{PP [\%]} & \textbf{HP [\%]}& \textbf{PP [\%]} & \textbf{HP [\%]} \\
		\midrule
        \multirow{3}{*}{AES} & LSQUIC & 16.83 & 0.31 & 14.83 & 0.08 \\
        & quiche & 15.30 & 2.45 & 14.75 & 0.92 \\
        & MsQuic & 16.79 & 0.68 & 27.25 & 0.66 \\
        \midrule
        \multirow{3}{*}{ChaCha20} & LSQUIC & 21.68 & 3.29 & 20.40 & 2.71 \\
        & quiche & 21.02 & 4.49 & 20.37 & 3.32 \\
        & MsQuic & 29.12 & 3.69 & 43.32 & 5.23 \\
        \midrule
        \multirow{3}{*}{NOOP} & LSQUIC & 3.11 & 0.03 & 2.96 & 0.01 \\
         & quiche & 1.91 & 2.07 & 1.35 & 0.07 \\
         & MsQuic & 1.94 & 0.59 & 2.54 & 0.01 \\
        \midrule
        NOENC & MsQuic & 0.37 & 0.52 & 0.05 & 0.01 \\
	    \bottomrule
	\end{tabular}
\end{table}

The results show that the packet protection is the primary contributor to the \ac{cpu} time consumption on both endpoints.
As expected, the results with an \ac{aes} cipher show a lower percentage of samples for packet and header protection than those with a ChaCha20 cipher.
The fact that ChaCha20 does not profit from hardware acceleration is reflected in the higher percentage of samples for packet and header protection and a lower goodput, as it was shown in \Cref{fig:ciphers}.

It is also noticeable that the percentage of samples for header protection is higher for \quiche, also with the \noopcipher.
This is caused by the fact that \quiche performs operations for header parsing in the same function where the header protection is removed.
After a closer look at the \quiche source code, the operations with zero-copy mutable byte buffers and the used return type are the main contributors to the higher cost for header protection.

When analyzing the results with the \noopcipher, it can be seen that these operations contribute around \sperc{2} on the \quiche client.
After subtracting the \sperc{2} for header protection, from the other values for \quiche clients, we receive similar values as for the other clients.

Comparing the results for \msquic with \noenc to the results with the \noopcipher, we can see a further reduction of the collected samples for packet protection.
This computational difference is the cost of the \noopcipher, which still calls \textit{OpenSSL} then executing \texttt{memcpy()}.

As we have already shown in \Cref{sec:hp-perf}, the header protection for \ac{aes} ciphers does not have a significant impact on the performance, contributing less than \sperc{1} to the total \ac{cpu} time consumption on both endpoints.

\subsection{MTU Impact}

Even though transmitting IP packets of larger than \SI{1500}{\byte} through the Internet is unrealistic, it is attractive for datacenter and local company networks.
As with \ac{tcp}, the \ac{mtu} has a significant impact on the performance of \quic.
With larger \acp{mtu}, the amount of packets to be sent is reduced, resulting in fewer per-packet overheads.
To elaborate on the advantages of larger \acp{mtu} in combination with \quic's always-on encryption, we performed measurements with \ac{mtu} values of \SI{1500}{\byte}, \SI{3000}{\byte}, and \SI{6000}{\byte}.
For each \ac{mtu}, we measured with \lsquic, \msquic and the \tcptls stack for comparison, each one with \ac{aes} and \noopcipher.
It has to be noted that \tcptls stack is limited by the link rate of \SI{10}{\Gbps} and therefore does not benefit from larger \acp{mtu} here.
The base \ac{mtu} was set to \SI{1500}{\byte} for measurements with a \ac{mtu} over \SI{1500}{\byte}.
The implementations perform \ac{mtu} discovery and adjust the \ac{mtu} accordingly.
We do not include results with \quiche here, as \quiche does not support changing the \ac{mtu} without introducing major changes to the library.

\begin{figure}
    \centering
    \includegraphics[width=\linewidth]{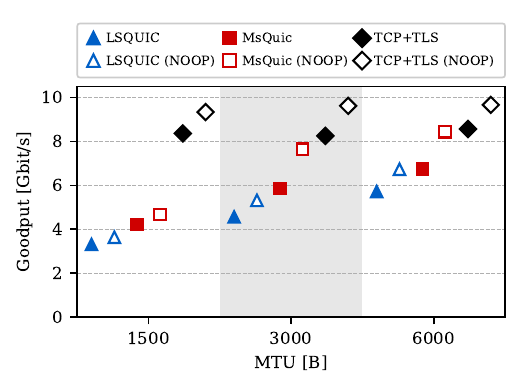}
    \caption{Goodput achieved with different MTUs for different \quic implementations, each one with \ac{aes} and \noopcipher.}%
    \label{fig:mtu}
\end{figure}

In \Cref{fig:mtu}, the results of the measurements are shown.
As expected, the goodput increases with larger \acp{mtu} for every tested implementation.
Both \lsquic and \msquic show an increase in goodput of almost \sperc{40} when the \ac{mtu} is increased from \SI{1500}{\byte} to \SI{3000}{\byte}.

When looking at packets sent from server to client,
both \quic implementations reach an average packet size of at least \SI{1498}{\byte} when the \ac{mtu} is set to \SI{1500}{\byte}.
\tcptls only reaches such values with our \noopcipher, sending packets with a slightly lower average size of \SI{1489}{\byte} with encryption enabled.
For the increased \ac{mtu} of \SI{3000}{\byte}, \msquic and \tcptls reach an average frame size between \SI{2953}{\byte} and \SI{2999}{\byte}.
\lsquic does not use the larger \ac{mtu} and only reaches an average frame size of around \SI{2350}{\byte}, leaving more than \SI{600}{\byte} per packet unused.
With the \ac{mtu} set to \SI{6000}{\byte}, this behavior is even more pronounced.
The \lsquic server sends packets of an average size of \SI{3510}{\byte},
while the \msquic makes use of the larger \ac{mtu} and sends packets of an average size of more than \SI{5960}{\byte}.
However, for none of the tested \quic implementations, the relative gain in goodput when switching to the \noopcipher increases with larger \acp{mtu}.

As the goodput gap between \quic and \tcptls decreases with larger \acp{mtu}, the performance of \quic is more competitive with \tcptls on \SI{10}{\Gbps} links.
In controlled environments with high bandwidth scenarios like datacenters, increasing the \ac{mtu} makes \quic a viable alternative to \tcptls.

\begin{table}[bt]
  \newcommand{\tablextwofivefiveonenine}{\textbf{\xtwofivefiveonenine}}
  \newcommand{\tablebikelone}{\bikelone}
  \newcommand{\tablekyberfivehundredtwelve}{\kyberfivehundredtwelve}
  \newcommand{\tablehqconehundredtwentyeight}{\hqconehundredtwentyeight}
  \newcommand{\tableptwohundredfiftysix}{\textbf{\ptwohundredfiftysix}}
  \newcommand{\tableptwohundredfiftysixbikelone}{\ptwohundredfiftysix + \bikelone}
  \newcommand{\tableptwohundredfiftysixkyberfivehundredtwelve}{\ptwohundredfiftysix + \kyberfivehundredtwelve}
  \newcommand{\tableptwohundredfiftysixhqconehundredtwentyeight}{\ptwohundredfiftysix + \hqconehundredtwentyeight}
  \newcommand{\tablebikelthree}{\bikelthree}
  \newcommand{\tablekybersevenhundredsixtyeight}{\kybersevenhundredsixtyeight}
  \newcommand{\tablehqconehundredninetytwo}{\hqconehundredninetytwo}
  \newcommand{\tablepthreehundredeightyfour}{\textbf{\pthreehundredeightyfour}}
  \newcommand{\tablepthreehundredeightyfourbikelthree}{\pthreehundredeightyfour + \bikelthree}
  \newcommand{\tablepthreehundredeightyfourkybersevenhundredsixtyeight}{\pthreehundredeightyfour + \kybersevenhundredsixtyeight}
  \newcommand{\tablepthreehundredeightyfourhqconehundredninetytwo}{\pthreehundredeightyfour + \hqconehundredninetytwo}
  \newcommand{\tablebikelfive}{\bikelfive}
  \newcommand{\tablekyberonethousandtwentyfour}{\kyberonethousandtwentyfour}
  \newcommand{\tablehqctwohundredfiftysix}{\hqctwohundredfiftysix}
  \newcommand{\tablepfivehundredtwentyone}{\textbf{\pfivehundredtwentyone}}
  \newcommand{\tablepfivehundredtwentyonebikelfive}{\pfivehundredtwentyone + \bikelfive}

  \newcommand{\tablepfivehundredtwentyonekyberonethousandtwentyfour}{\pfivehundredtwentyone + \kyberonethousandtwentyfour}
  \newcommand{\tablepfivehundredtwentyonehqctwohundredfiftysix}{\pfivehundredtwentyone + \hqctwohundredfiftysix}
  \footnotesize
  \centering
    \caption[Post-quantum \acsp{kem} Measurement Results]{\label{tab:evaluation_pqc_kems}Median \acs{ttfb} and \quic handshake packet count for different \quic implementations and post-quantum \acsp{kem} at different \acs{nist} levels. In the packet count columns, the first/second number represents the amount of packets sent by the client/server. All instantiations were measured with an \rsatwothousandfortyeight certificate and \aeadaesonehundredtwentyeightgcm as \acs{aead} algorithm. \textbf{Bold} algorithms are not quantum-safe.}
    \begin{tabularx}{\linewidth}{llrrrr}
      \toprule
       & \multirow[b]{2}{*}{\acs{kem}} & \multicolumn{2}{c}{\acs{ttfb} [\si{\milli\second}]} & \multicolumn{2}{c}{Packets [C / S]} \\
      \cmidrule(rl){3-4} \cmidrule(rl){5-6}
       & &\multicolumn{1}{c}{\lsquic} & \multicolumn{1}{c}{\quiche} & \multicolumn{1}{c}{\lsquic} & \multicolumn{1}{c}{\quiche} \\
      \midrule
      \nistlone \begin{filecontents*}{data-combined-kems-l1.csv}
display_name;lsquic_client_handshake_done;lsquic_client_handshake_packets;lsquic_server_handshake_packets;quiche_client_handshake_done;quiche_client_handshake_packets;quiche_server_handshake_packets;pub_key_size;secret_key_size;ciphertext_size
\tablextwofivefiveonenine;3.91;3;3;3.57;3;3;32;32;32
\tablekyberfivehundredtwelve;4.08;3;3;3.39;3;3;800;1632;768
\tablebikelone;6.59;4;4;5.86;4;5;1541;5223;1573
\tablehqconehundredtwentyeight;5.57;5;6;4.21;5;8;2249;2289;4481
\tableptwohundredfiftysix;3.90;3;3;3.49;3;3;65;32;65
\tableptwohundredfiftysixkyberfivehundredtwelve;4.43;3;3;3.74;3;3;865;1664;833
\tableptwohundredfiftysixbikelone;6.95;4;4;6.27;4;5;1606;5255;1638
\tableptwohundredfiftysixhqconehundredtwentyeight;5.99;5;6;4.52;5;8;2314;2321;4546
\end{filecontents*}
\csvreader[head to column names, separator=semicolon, late after line = \\]{data-combined-kems-l1.csv}{%
      display_name=\one, lsquic_client_handshake_done=\two, quiche_client_handshake_done=\three, lsquic_client_handshake_packets=\fourl, lsquic_server_handshake_packets=\fivel, quiche_client_handshake_packets=\fourq, quiche_server_handshake_packets=\fiveq
      }{
        & \one & \two & \three & \fourl~/~\fivel & \fourq~/~\fiveq
      }
      \midrule
      \nistlthree \begin{filecontents*}{data-combined-kems-l3.csv}
display_name;lsquic_client_handshake_done;lsquic_client_handshake_packets;lsquic_server_handshake_packets;quiche_client_handshake_done;quiche_client_handshake_packets;quiche_server_handshake_packets;pub_key_size;secret_key_size;ciphertext_size
\tablekybersevenhundredsixtyeight;4.23;4;3;3.78;4;5;1184;2400;1088
\tablebikelthree;11.75;5;5;10.49;5;6;3083;10105;3115
\tablehqconehundredninetytwo;7.57;6;10;4.81;7;12;4522;4562;9026
\tablepthreehundredeightyfour;7.36;3;3;6.76;3;3;97;48;97
\tablepthreehundredeightyfourkybersevenhundredsixtyeight;8.99;4;4;8.67;4;5;1281;2448;1185
\tablepthreehundredeightyfourbikelthree;16.69;5;5;15.14;6;7;3180;10153;3212
\tablepthreehundredeightyfourhqconehundredninetytwo;12.42;7;10;9.44;7;12;4619;4610;9123
\end{filecontents*}
\csvreader[head to column names, separator=semicolon, late after line = \\]{data-combined-kems-l3.csv}{%
      display_name=\one, lsquic_client_handshake_done=\two, quiche_client_handshake_done=\three, lsquic_client_handshake_packets=\fourl, lsquic_server_handshake_packets=\fivel, quiche_client_handshake_packets=\fourq, quiche_server_handshake_packets=\fiveq
      }{
        & \one & \two & \three & \fourl~/~\fivel & \fourq~/~\fiveq
      }
      \midrule
      \nistlfive \begin{filecontents*}{data-combined-kems-l5.csv}
display_name;lsquic_client_handshake_done;lsquic_client_handshake_packets;lsquic_server_handshake_packets;quiche_client_handshake_done;quiche_client_handshake_packets;quiche_server_handshake_packets;pub_key_size;secret_key_size;ciphertext_size
\tablekyberonethousandtwentyfour;4.43;4;4;3.81;4;5;1568;3168;1568
\tablebikelfive;22.27;7;7;20.08;7;8;5122;16494;5154
\tablehqctwohundredfiftysix;10.15;9;15;6.04;10;17;7245;7285;14469
\tablepfivehundredtwentyone;12.24;3;3;11.86;3;3;133;66;133
\tablepfivehundredtwentyonekyberonethousandtwentyfour;15.98;4;4;15.12;4;5;1701;3234;1701
\tablepfivehundredtwentyonebikelfive;33.67;7;7;31.99;7;8;5255;16560;5287
\tablepfivehundredtwentyonehqctwohundredfiftysix;22.11;9;15;17.78;10;17;7378;7351;14602
\end{filecontents*}
\csvreader[head to column names, separator=semicolon, late after line = \\]{data-combined-kems-l5.csv}{%
      display_name=\one, lsquic_client_handshake_done=\two, quiche_client_handshake_done=\three, lsquic_client_handshake_packets=\fourl, lsquic_server_handshake_packets=\fivel, quiche_client_handshake_packets=\fourq, quiche_server_handshake_packets=\fiveq
      }{
        & \one & \two & \three & \fourl~/~\fivel & \fourq~/~\fiveq
      }
      \bottomrule
    \end{tabularx}
\end{table}

\subsection{Post-Quantum Cryptography}

With the \boringssl fork from \ac{oqs} introduced in \Cref{sec:approach-ssl}, we measured the additional costs by using \ac{pqc} during the \quic handshake.
\msquic was not included in these measurements, as it uses \openssl.
Since only the handshake is affected in these measurements the goodput is not a suitable metric.
The filesize of the requested file was reduced to \SI{1}{\byte} and the time between the different steps of the handshake was measured.
We define the metric \ac{ttfb} as the time between the client sending its \clienthello and being able to send its \ac{http}/3 request to the server.
The amount of \acp{rtt} needed for the handshake can be neglected here, as the \ac{rtt} in our measurement setup is below \SI{0.1}{\milli\second}.
Additionally, the difference in \acp{rtt} until the client is able to start sending the \ac{http}/3 request is not greater than 1 \ac{rtt} between \lsquic and \quiche for most of the performed measurements.

First, we take a look at the \acp{kem} \kyber, \bike, and HQC.
\kyber was chosen for standardization by the \ac{nist}~\cite{fips203} while BIKE and HQC were selected for the fourth round of the \ac{nist} \ac{pqc} standardization process~\cite{nist-pqc-status-report-third-round}.

As a baseline, the traditional key exchange method ECDHE was tested with \curvetwofivefiveonenine (\xtwofivefiveonenine), \ptwohundredfiftysix, \pthreehundredeightyfour, and \pfivehundredtwentyone.
Additionally, hybrid \ac{kem} algorithms were measured, which combine the respective post-quantum \ac{kem} with pre-quantum ECDHE using \ptwohundredfiftysix, \pthreehundredeightyfour or \pfivehundredtwentyone.
By using multiple key exchange algorithms simultaneously and combining the results, security is provided even if one of the two used algorithms turns out to be broken~\cite{ietf-tls-hybrid-design-09}.
This might happen if either the new quantum-safe algorithms turn out to be insecure or if the traditional algorithms are broken by a quantum computer.

\begin{figure}
    \centering
    \includegraphics[width=\linewidth]{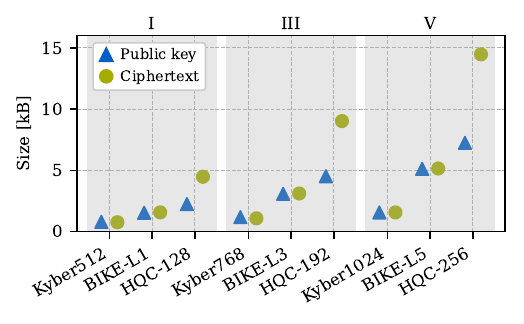}
    \caption{Public key and ciphertext sizes of the benchmarked post-quantum \acp{kem} grouped by \ac{nist} level.}%
    \label{fig:pqc_sizes}
\end{figure}

In \Cref{tab:evaluation_pqc_kems}, the results of the measurements are shown.
The tested \acp{kem} are grouped by their respective \ac{nist} levels.
In all \ac{nist} levels, \kyber shows the fastest handshake from the tested post-quantum \acp{kem}.
As \kyber is the only \ac{kem} that is lattice-based, it profits from smaller key sizes.
The public key sent as key share in the \clienthello and the ciphertext sent as a key share in the \serverhello influence the number of packets each endpoint sends during the handshake.
The different sizes of the public keys and ciphertexts for the benchmarked \ac{pqc} \ac{kem} algorithms are visualized in \Cref{fig:pqc_sizes}.
For all tested ECDHE algorithms, the size of the public keys and ciphertexts is between \SI{32}{\byte} and \SI{133}{\byte} and therefore negligible small.
\kyber provides pleasantly small public keys and ciphertexts of under \SI{2}{\kilo\byte}, even with \kyberonethousandtwentyfour on \ac{nist} level \nistlfive.
While a \SI{800}{\byte} public key is sent in the \clienthello with \kyberfivehundredtwelve, \hqconehundredtwentyeight needs to send \SI{2249}{\byte}.
The \ac{ttfb} measured with \kyber as \ac{kem} is only slightly increased compared to \xtwofivefiveonenine, even on \ac{nist} level \nistlfive.

We noticed a slightly higher \ac{ttfb} with \lsquic than with \quiche.
After taking a closer look, it was noticeable that the \lsquic client needs more time to process the \serverhello message.
With \hqctwohundredfiftysix, this difference is most pronounced, with the \lsquic client needing around \SI{5}{\milli\second} instead of the around \SI{1.5}{\milli\second} \quiche needs.

In our measurements, \bike is the slowest \ac{kem} on all \ac{nist} levels.
With \bikelfive, the \ac{ttfb} extends five times that of \kyberonethousandtwentyfour and more than twice that of the likewise code-based \hqctwohundredfiftysix.
Although \hqctwohundredfiftysix has nearly triple the ciphertext size of \bikelfive and a bigger public key, it still has an over \qty{50}{\percent} lower \ac{ttfb}.

\begin{table*}[bt]
    \newcommand{\tablersaonethousandtwentyfour}{\textbf{\rsaonethousandtwentyfour}}
    \newcommand{\tablersatwothousandfortyeight}{\textbf{\rsatwothousandfortyeight (default)}}
    \newcommand{\tablersathreethousandseventytwo}{\textbf{\rsathreethousandseventytwo}}
    \newcommand{\tablersafourthousandninetysix}{\textbf{\rsafourthousandninetysix}}
    \newcommand{\tablefalconfivehundredtwelve}{\falconfivehundredtwelve}
    \newcommand{\tablesphincsshatwoonehundredtwentyeightfsimple}{\sphincsshatwoonehundredtwentyeightf}
    \newcommand{\tablesphincsshatwoonehundredtwentyeightssimple}{\sphincsshatwoonehundredtwentyeights}
    \newcommand{\tablesphincsshakeonehundredtwentyeightfsimple}{\sphincsshakeonehundredtwentyeightf}
    \newcommand{\tablesphincsshakeonehundredtwentyeightssimple}{\sphincsshakeonehundredtwentyeights}
    \newcommand{\tabledilithiumthree}{\dilithiumthree}
    \newcommand{\tablesphincsshatwoonehundredninetytwofsimple}{\sphincsshatwoonehundredninetytwof}
    \newcommand{\tablesphincsshatwoonehundredninetytwossimple}{\sphincsshatwoonehundredninetytwos}
    \newcommand{\tablesphincsshakeonehundredninetytwofsimple}{\sphincsshakeonehundredninetytwof}
    \newcommand{\tablesphincsshakeonehundredninetytwossimple}{\sphincsshakeonehundredninetytwos}
    \newcommand{\tabledilithiumfive}{\dilithiumfive}
    \newcommand{\tablefalcononethousandtwentyfour}{\falcononethousandtwentyfour}
    \newcommand{\tablesphincsshatwotwohundredfiftysixfsimple}{\sphincsshatwotwohundredfiftysixf}
    \newcommand{\tablesphincsshatwotwohundredfiftysixssimple}{\sphincsshatwotwohundredfiftysixs}
    \newcommand{\tablesphincsshaketwohundredfiftysixfsimple}{\sphincsshaketwohundredfiftysixf}
    \newcommand{\tablesphincsshaketwohundredfiftysixssimple}{\sphincsshaketwohundredfiftysixs}
    \footnotesize
    \centering
    \caption[Post-quantum Signature Schemes Measurement Results]{\label{tab:evaluation_pqc_signature_schemes}Median \acs{ttfb}, \quic handshake packet count, public key, signature, and certificate sizes for different \quic implementations and post-quantum signature algorithms at different \acs{nist} levels. In the packet count columns, the first/second number represents the amount of packets sent by the client/server. All instantiations were measured with \xtwofivefiveonenine as \acs{kem} and \aeadaesonehundredtwentyeightgcm as \acs{aead} algorithm. \textbf{Bold} algorithms are not quantum-safe.}
    \begin{tabularx}{\linewidth}{lXrrrrrrr}

      \toprule
       & \multirow[b]{2}{*}{Signature algorithm} & \multicolumn{2}{c}{\acs{ttfb} [\si{\milli\second}]} & \multicolumn{2}{c}{Packets [C / S]} & \multicolumn{3}{c}{Sizes [\unit{\byte}]}\\
      \cmidrule(rl){3-4} \cmidrule(rl){5-6} \cmidrule(rl){7-9}
       & &\multicolumn{1}{c}{\lsquic} & \multicolumn{1}{c}{\quiche} & \multicolumn{1}{c}{\lsquic} & \multicolumn{1}{c}{\quiche} & \multicolumn{1}{c}{Public key} & \multicolumn{1}{c}{Signature} & \multicolumn{1}{c}{Certificate}\\
      \midrule

      \begin{filecontents*}{data-combined-sigs-l0.csv}
display_name;lsquic_client_handshake_done;lsquic_client_handshake_packets;lsquic_server_handshake_packets;quiche_client_handshake_done;quiche_client_handshake_packets;quiche_server_handshake_packets;pub_key_size;sig_size;cert_size
\tablersaonethousandtwentyfour;2.11;3;2;1.88;3;2;128;128;528
\tablersatwothousandfortyeight;3.91;3;3;3.57;3;3;256;256;789
\end{filecontents*}
\csvreader[head to column names, separator=semicolon, late after line = \\]{data-combined-sigs-l0.csv}{%
        display_name=\one, lsquic_client_handshake_done=\two, quiche_client_handshake_done=\three, lsquic_client_handshake_packets=\fourl, lsquic_server_handshake_packets=\fivel, quiche_client_handshake_packets=\fourq, quiche_server_handshake_packets=\fiveq, pub_key_size=\six, sig_size=\seven, cert_size=\eight
      }{
       & \one & \two & \three & \fourl~/~\fivel & \fourq~/~\fiveq & \six & \seven & \eight
      }
      \midrule
      \nistlone \begin{filecontents*}{data-combined-sigs-l1.csv}
display_name;lsquic_client_handshake_done;lsquic_client_handshake_packets;lsquic_server_handshake_packets;quiche_client_handshake_done;quiche_client_handshake_packets;quiche_server_handshake_packets;pub_key_size;sig_size;cert_size
\tablersafourthousandninetysix;14.02;3;3;13.46;3;3;512;512;1301
\tablefalconfivehundredtwelve;2.78;3;4;2.05;3;4;897;666;1793
\tablesphincsshatwoonehundredtwentyeightssimple;198.46;4;16;195.00;5;17;32;7856;8131
\tablesphincsshatwoonehundredtwentyeightfsimple;28.54;4;32;20.89;5;33;32;17088;17363
\tablesphincsshakeonehundredtwentyeightssimple;385.60;5;16;367.45;5;17;32;7856;8131
\tablesphincsshakeonehundredtwentyeightfsimple;50.95;4;32;41.89;5;33;32;17088;17363
\end{filecontents*}
\csvreader[head to column names, separator=semicolon, late after line = \\]{data-combined-sigs-l1.csv}{%
        display_name=\one, lsquic_client_handshake_done=\two, quiche_client_handshake_done=\three, lsquic_client_handshake_packets=\fourl, lsquic_server_handshake_packets=\fivel, quiche_client_handshake_packets=\fourq, quiche_server_handshake_packets=\fiveq, pub_key_size=\six, sig_size=\seven, cert_size=\eight
      }{
       & \one & \two & \three & \fourl~/~\fivel & \fourq~/~\fiveq & \six & \seven & \eight
      }
      \midrule
      \nistlthree \begin{filecontents*}{data-combined-sigs-l3.csv}
display_name;lsquic_client_handshake_done;lsquic_client_handshake_packets;lsquic_server_handshake_packets;quiche_client_handshake_done;quiche_client_handshake_packets;quiche_server_handshake_packets;pub_key_size;sig_size;cert_size
\tabledilithiumthree;4.26;4;10;2.14;4;10;1952;3293;5508
\tablesphincsshatwoonehundredninetytwossimple;325.46;5;31;325.37;5;32;48;16224;16516
\tablesphincsshatwoonehundredninetytwofsimple;-;-;-;35.31;6;67;48;35664;35956
\tablesphincsshakeonehundredninetytwossimple;590.73;6;31;601.04;5;32;48;16224;16516
\tablesphincsshakeonehundredninetytwofsimple;-;-;-;67.51;6;66;48;35664;35955
\end{filecontents*}
\csvreader[head to column names, separator=semicolon, late after line = \\]{data-combined-sigs-l3.csv}{%
        display_name=\one, lsquic_client_handshake_done=\two, quiche_client_handshake_done=\three, lsquic_client_handshake_packets=\fourl, lsquic_server_handshake_packets=\fivel, quiche_client_handshake_packets=\fourq, quiche_server_handshake_packets=\fiveq, pub_key_size=\six, sig_size=\seven, cert_size=\eight
      }{
       & \one & \two & \three & \fourl~/~\fivel & \fourq~/~\fiveq & \six & \seven & \eight
      }
      \midrule
      \nistlfive \begin{filecontents*}{data-combined-sigs-l5.csv}
display_name;lsquic_client_handshake_done;lsquic_client_handshake_packets;lsquic_server_handshake_packets;quiche_client_handshake_done;quiche_client_handshake_packets;quiche_server_handshake_packets;pub_key_size;sig_size;cert_size
\tabledilithiumfive;5.15;4;13;2.30;4;12;2592;4595;7449
\tablefalcononethousandtwentyfour;4.25;4;7;3.02;4;6;1793;1280;3299
\tablesphincsshatwotwohundredfiftysixssimple;-;-;-;297.50;6;56;64;29792;30100
\tablesphincsshatwotwohundredfiftysixfsimple;-;-;-;68.21;7;91;64;49856;50164
\tablesphincsshaketwohundredfiftysixssimple;-;-;-;536.84;6;56;64;29792;30100
\tablesphincsshaketwohundredfiftysixfsimple;-;-;-;106.11;7;92;64;49856;50164
\end{filecontents*}
\csvreader[head to column names, separator=semicolon, late after line = \\]{data-combined-sigs-l5.csv}{%
        display_name=\one, lsquic_client_handshake_done=\two, quiche_client_handshake_done=\three, lsquic_client_handshake_packets=\fourl, lsquic_server_handshake_packets=\fivel, quiche_client_handshake_packets=\fourq, quiche_server_handshake_packets=\fiveq, pub_key_size=\six, sig_size=\seven, cert_size=\eight
      }{
       & \one & \two & \three & \fourl~/~\fivel & \fourq~/~\fiveq & \six & \seven & \eight
      }
      \bottomrule
    \end{tabularx}
  \end{table*}

As the hybrid \acp{kem} concatenate the public keys and ciphertexts of the post-quantum \ac{kem} and the pre-quantum ECDHE algorithm, the sizes for the hybrid \acp{kem} are the sum of the sizes of the \acp{kem} being hybridized.
With \pthreehundredeightyfour and \pfivehundredtwentyone, the \ac{ttfb} is approximately the sum of the \ac{ttfb} for \pthreehundredeightyfour, respectively, \pfivehundredtwentyone and the post-quantum \ac{kem}.
Due to the high cost of ECDHE in these cases, the hybrid \acp{kem} are more expensive than the respective post-quantum \acp{kem}.
The hybrid \acp{kem} with \ptwohundredfiftysix are only slightly slower than the used post-quantum \acp{kem} and therefore a good choice for a hybrid approach.

To summarize, the major bottleneck of post-quantum key exchange algorithms in our measurements is the larger amount of data that needs to be transferred during the handshake.
This increases the latency and \ac{ttfb}.
Since the sizes for hybrid schemes are only marginally larger and, with an efficient elliptic curve such as \ptwohundredfiftysix, only marginally slower, it is best to use them if post-quantum security is desired as they guarantee security even if the post-quantum \ac{kem} turns out to be insecure.

For post-quantum signature schemes, we selected \ac{FALCON}, \dilithium, and \sphincs, which are all chosen for standardization by \ac{nist}~\cite{nist-pqc-status-report-third-round}.
We fixed the key exchange algorithm to \xtwofivefiveonenine ECDHE for our measurements.
The traditional pre-quantum signature scheme \ac{RSA} is used as a baseline with \qty{1024}{\bit}, \qty{2048}{\bit}, and \qty{4096}{\bit} keys.

The results of the measurements are shown in \Cref{tab:evaluation_pqc_signature_schemes}, listing the aforementioned signature algorithms grouped by their respective \ac{nist} levels.
With increasing sizes of the public key and the signature, the certificate grows bigger.
Due to the increasing signature size, the \ac{tls} \tlscertificateverify also expands in size.
This leads to more packets being sent by the server, which can be observed with \dilithium, \ac{FALCON}, and \sphincs, where the server must send up to 92 packets in the handshake.
The client must also send extra packets to acknowledge the additional packets from the server.
\falconfivehundredtwelve is the post-quantum scheme with the slightest increase in latency: The \ac{ttfb} for client and server is only about \qty{0.6}{\milli\second} higher compared to \rsaonethousandtwentyfour.
\falconfivehundredtwelve even performed better, in terms of latency, than \rsatwothousandfortyeight.

For the \rsafourthousandninetysix, the \ac{ttfb} for the client and server rises to an extreme \qty{14}{\milli\second}, making it the slowest of the tested pre-quantum schemes.
It was slower than all the post-quantum schemes we looked at, except for \sphincs.
The \sphincs variants reach by far the highest \ac{ttfb} of all evaluated signature schemes, caused by the huge signature sizes.
The fast variants (denoted by an f suffix) are still slower than any other signature scheme measured.
The more compact signatures of the small variants (denoted by an s suffix) come at the expense of calculation time:
The \ac{ttfb} of \sphincsshakeonehundredninetytwos is over 142 times higher than \dilithiumthree's.
The versions of \sphincs that used \shake were about twice as slow as those using \shatwo, even though they produced signatures of the same size.

In conclusion, post-quantum signature schemes come with larger public key and signature sizes than the \ac{RSA} variants.
\ac{FALCON} is the quantum-safe signature scheme with the smallest signature and certificate size.
\dilithium is larger while still having an acceptable signature and certificate size compared to the hash-based \sphincs with huge signatures of up to \SI{49}{\kilo\byte}.
Moreover, we saw that increasing the \ac{RSA} key size is insufficient to improve security while keeping the performance impact minimal.
Instead, a performant post-quantum signature scheme like \ac{FALCON} should be employed if quantum security is desired.

As can be seen from the missing values for \lsquic in \Cref{tab:evaluation_pqc_signature_schemes}, the measurements with certificates larger than \SI{30}{\kilo\byte} have failed.
\lsquic's server had problems sending out the \serverhello.
For the handshake, \lsquic uses so-called \textit{mini connections}, which allocate less memory to protect the server from DoS attacks.
Those mini connections use bitmasks to keep track of packet numbers.\footnote{\url{https://lsquic.readthedocs.io/en/v4.0.0/internals.html\#mini-ietf-connection}}
Due to the variable length of the bitmasks, only up to 64 packets are supported.
This limit is exceeded with the huge signature and certificate sizes of \sphincs, as it can be seen for the \quiche measurements.
This indicates that \lsquic is not ready for post-quantum signature schemes with huge certificate and signature sizes like some parameter sets of \sphincs, even if its use in \quic is of questionable value because of the poor performance.

Another issue arises with large certificate sizes in combination with \quic's address validation.
The server is not allowed to send more than three times as many bytes as the number of received bytes if the client address is not yet validated~\cite[Section 8.1]{rfc9000}.
To validate the client's address before completing the \ac{tls} handshake, the server can send a \retrypacket packet.
However, this causes an additional \ac{rtt} and therefore increases the \ac{ttfb}.
By using larger \ac{mtu} values, the \clienthello can be padded to larger sizes, which can mitigate this issue.

\section{Conclusion}
\label{sec:conclusion}

In this work, we evaluate the impact of cryptography on \quic performance.
The cost of symmetric cryptography during the connection, consisting of packet and header protection, is analyzed.
We additionally evaluate the asymmetric cryptography happening during the handshake with precise measurements.
We integrate quantum-safe cryptographic algorithms into the chosen \quic implementations to evaluate the performance impact and identify possible problems when integrating \ac{pqc} into \quic.

Our analysis of cipher suites shows that hardware-accelerated \texttt{AEAD\_AES\_128\_GCM} is the most efficient \ac{aead} algorithm for header and packet protection.
Compared to packet protection, header protection has little impact on \ac{cpu} time consumption and goodput.
Especially for \ac{aes} ciphers, the header protection is virtually free.
We reveal that increasing the \ac{mtu} does not mitigate the impact of encryption, as using the \noop cipher shows performance improvements also for larger \acp{mtu}.

The integration of quantum-safe cryptographic algorithms into \quic is feasible without major changes to the \quic libraries using \boringssl.
While algorithms with larger key sizes or more complex calculations have a significant impact on the handshake duration, algorithms like \kyber and \dilithium are promising candidates for post-quantum secure \quic, as they have a low impact on the handshake duration.
Large certificate sizes lead to different problems in our measurements, as the packet number space for \handshakepacket packets might be limited or \quic's address validation mechanism can cause an extra \acp{rtt}.

To allow for evaluations of other \quic implementations, we publish the modified \boringssl and \openssl libraries~\cite{tls-lib-publication}.

\section*{Acknowledgment}
\label{sec:acknowledgment}

The European Union's Horizon 2020 research and innovation programme funded this work under grant agreements No 101008468 and 101079774.
Additionally, we received funding by the Bavarian Ministry of Economic Affairs, Regional Development and Energy as part of the project 6G Future Lab Bavaria.
This work is partially funded by Germany Federal Ministry of Education and Research (BMBF) under the projects 6G-life (16KISK001K) and 6G-ANNA (16KISK107) and the German Research Foundation under the project HyperNIC (CA595/13-1).

\bibliographystyle{IEEEtranN}
{
\footnotesize
\bibliography{IEEEabrv,paper_strip}

\begin{thebibliography}{37}
\providecommand{\natexlab}[1]{#1}
\providecommand{\url}[1]{#1}
\csname url@samestyle\endcsname
\providecommand{\newblock}{\relax}
\providecommand{\bibinfo}[2]{#2}
\providecommand{\BIBentrySTDinterwordspacing}{\spaceskip=0pt\relax}
\providecommand{\BIBentryALTinterwordstretchfactor}{4}
\providecommand{\BIBentryALTinterwordspacing}{\spaceskip=\fontdimen2\font plus
\BIBentryALTinterwordstretchfactor\fontdimen3\font minus
  \fontdimen4\font\relax}
\providecommand{\BIBforeignlanguage}[2]{{%
\expandafter\ifx\csname l@#1\endcsname\relax
\typeout{** WARNING: IEEEtranN.bst: No hyphenation pattern has been}%
\typeout{** loaded for the language `#1'. Using the pattern for}%
\typeout{** the default language instead.}%
\else
\language=\csname l@#1\endcsname
\fi
#2}}
\providecommand{\BIBdecl}{\relax}
\BIBdecl

\bibitem[Iyengar and Thomson(2021)]{rfc9000}
\BIBentryALTinterwordspacing
J.~Iyengar and M.~Thomson, ``{QUIC: A UDP-Based Multiplexed and Secure
  Transport},'' RFC 9000, May 2021. [Online]. Available:
  \url{https://rfc-editor.org/rfc/rfc9000.txt}
\BIBentrySTDinterwordspacing

\bibitem[Thomson and Turner(2021)]{rfc9001}
\BIBentryALTinterwordspacing
M.~Thomson and S.~Turner, ``{Using TLS to Secure QUIC},'' RFC 9001, May 2021.
  [Online]. Available: \url{https://rfc-editor.org/rfc/rfc9001.txt}
\BIBentrySTDinterwordspacing

\bibitem[{QUIC IETF Mailinglist}(2020)]{quic-crypto-discussion}
\BIBentryALTinterwordspacing
{QUIC IETF Mailinglist}. (2020) {A non-TLS standard is needed}. Accessed:
  2024-02-29. [Online]. Available:
  \url{https://mailarchive.ietf.org/arch/msg/quic/SBetxLwCq5I7un2tkzFb7tXhJMU/}
\BIBentrySTDinterwordspacing

\bibitem[{QUIC IETF Mailinglist}(2024)]{quic-crypto-discussion-2}
\BIBentryALTinterwordspacing
------. (2024) {Historic TLS Discussion}. Accessed: 2024-03-08. [Online].
  Available:
  \url{https://mailarchive.ietf.org/arch/msg/quic/rDUtUDVqz95JspgptALSNYcnn5c/}
\BIBentrySTDinterwordspacing

\bibitem[Jaeger et~al.(2023)Jaeger, Zirngibl, Kempf, Ploch, and
  Carle]{jaeger2023quic}
B.~Jaeger, J.~Zirngibl, M.~Kempf, K.~Ploch, and G.~Carle, ``{QUIC on the
  Highway: Evaluating Performance on High-Rate Links},'' in \emph{International
  Federation for Information Processing (IFIP) Networking 2023 Conference (IFIP
  Networking 2023)}, Barcelona, Spain, Jun. 2023.

\bibitem[{United States National Institute of Standards and
  Technology}(2023)]{pqc-standardization}
\BIBentryALTinterwordspacing
{United States National Institute of Standards and Technology}, ``{Post-Quantum
  Cryptography Standardization},'' 2023, accessed: 2024-02-29. [Online].
  Available: \url{https://csrc.nist.gov/projects/post-quantum-cryptography}
\BIBentrySTDinterwordspacing

\bibitem[Rescorla(2018)]{rfc8446}
\BIBentryALTinterwordspacing
E.~Rescorla, ``{The Transport Layer Security (TLS) Protocol Version 1.3},'' RFC
  8446, Aug. 2018. [Online]. Available:
  \url{https://www.rfc-editor.org/info/rfc8446}
\BIBentrySTDinterwordspacing

\bibitem[Gueron(2010)]{intelaesni}
\BIBentryALTinterwordspacing
S.~Gueron. (2010) {Intel® Advanced Encryption Standard (AES) New Instructions
  Set}. Accessed: 2024-02-29. [Online]. Available:
  \url{https://www.intel.com/content/dam/doc/white-paper/advanced-encryption-standard-new-instructions-set-paper.pdf}
\BIBentrySTDinterwordspacing

\bibitem[Nir and Langley(2015)]{rfc7539}
\BIBentryALTinterwordspacing
Y.~Nir and A.~Langley, ``{ChaCha20 and Poly1305 for IETF Protocols},'' RFC
  7539, May 2015. [Online]. Available:
  \url{https://www.rfc-editor.org/info/rfc7539}
\BIBentrySTDinterwordspacing

\bibitem[Shor(1999)]{shor1999polynomial}
P.~W. Shor, ``{Polynomial-Time Algorithms for Prime Factorization and Discrete
  Logarithms on a Quantum Computer},'' \emph{SIAM review}, 1999.

\bibitem[Grover(1996)]{10.1145/237814.237866}
L.~K. Grover, ``{A Fast Quantum Mechanical Algorithm for Database Search},'' in
  \emph{Proceedings of the Twenty-Eighth Annual ACM Symposium on Theory of
  Computing}, 1996.

\bibitem[{Open Quantum Safe project}(2024{\natexlab{a}})]{oqs-algorithms}
\BIBentryALTinterwordspacing
{Open Quantum Safe project}, ``{Algorithms in liboqs},'' 2024, accessed:
  2024-02-29. [Online]. Available:
  \url{https://openquantumsafe.org/liboqs/algorithms/}
\BIBentrySTDinterwordspacing

\bibitem[{Bas Westerbaan}(March 5, 2024)]{cf-pq-blog-2024}
\BIBentryALTinterwordspacing
{Bas Westerbaan}, ``{The state of the post-quantum Internet},'' March 5, 2024,
  accessed: 2024-03-08. [Online]. Available:
  \url{https://blog.cloudflare.com/pq-2024}
\BIBentrySTDinterwordspacing

\bibitem[Bernstein et~al.(2008)Bernstein, Buchmann, and
  Dahmen]{10.5555/1522375}
D.~J. Bernstein, J.~Buchmann, and E.~Dahmen, \emph{{Post Quantum
  Cryptography}}, 1st~ed.\hskip 1em plus 0.5em minus 0.4em\relax Springer
  Publishing Company, Incorporated, 2008.

\bibitem[{LiteSpeed Tech}(2024)]{lsquic}
\BIBentryALTinterwordspacing
{LiteSpeed Tech}. (2024) {lsquic}. Accessed: 2024-02-13. [Online]. Available:
  \url{https://github.com/litespeedtech/lsquic}
\BIBentrySTDinterwordspacing

\bibitem[{Cloudflare}(2024)]{quiche}
\BIBentryALTinterwordspacing
{Cloudflare}. (2024) {quiche}. Accessed: 2024-02-13. [Online]. Available:
  \url{https://github.com/cloudflare/quiche}
\BIBentrySTDinterwordspacing

\bibitem[{Microsoft}(2024)]{msquic}
\BIBentryALTinterwordspacing
{Microsoft}. (2024) {MsQuic}. Accessed: 2024-02-13. [Online]. Available:
  \url{https://github.com/microsoft/msquic}
\BIBentrySTDinterwordspacing

\bibitem[Yang et~al.(2020)Yang, Eggert, Ott, Uhlig, Sun, and
  Antichi]{yang2020quicnicoffloading}
\BIBentryALTinterwordspacing
X.~Yang, L.~Eggert, J.~Ott, S.~Uhlig, Z.~Sun, and G.~Antichi, ``{Making QUIC
  Quicker With NIC Offload},'' in \emph{Proceedings of the Workshop on the
  Evolution, Performance, and Interoperability of QUIC}, 2020. [Online].
  Available: \url{https://doi.org/10.1145/3405796.3405827}
\BIBentrySTDinterwordspacing

\bibitem[{NetApp}(2024)]{quant}
\BIBentryALTinterwordspacing
{NetApp}. (2024) {Quant}. Accessed: 2024-02-13. [Online]. Available:
  \url{https://github.com/NTAP/quant}
\BIBentrySTDinterwordspacing

\bibitem[{H2O Project}(2024)]{quicly}
\BIBentryALTinterwordspacing
{H2O Project}. (2024) {Quicly}. Accessed: 2024-02-13. [Online]. Available:
  \url{https://github.com/h2o/quicly}
\BIBentrySTDinterwordspacing

\bibitem[{Private Octopus}(2024)]{picoquic}
\BIBentryALTinterwordspacing
{Private Octopus}. (2024) {picoquic}. Accessed: 2024-02-13. [Online].
  Available: \url{https://github.com/private-octopus/picoquic}
\BIBentrySTDinterwordspacing

\bibitem[{Facebook}(2024)]{mvfst}
\BIBentryALTinterwordspacing
{Facebook}. (2024) {mvfst}. Accessed: 2024-02-13. [Online]. Available:
  \url{https://github.com/facebookincubator/mvfst}
\BIBentrySTDinterwordspacing

\bibitem[Bauer et~al.(2023)Bauer, Sattler, Zirngibl, Schwarzenberg, and
  Carle]{bauer2023tcpandquic}
S.~Bauer, P.~Sattler, J.~Zirngibl, C.~Schwarzenberg, and G.~Carle,
  ``{Evaluating the Benefits: Quantifying the Effects of TCP Options, QUIC, and
  CDNs on Throughput},'' in \emph{Proceedings of the Applied Networking
  Research Workshop}, 2023.

\bibitem[Yu and Benson(2021)]{yu2021quicproductionperformance}
A.~Yu and T.~A. Benson, ``{Dissecting Performance of Production QUIC},'' in
  \emph{Proceedings of the Web Conference 2021}, 2021.

\bibitem[Marx et~al.(2020)Marx, Herbots, Lamotte, and
  Quax]{marx2020implementationdiversity}
R.~Marx, J.~Herbots, W.~Lamotte, and P.~Quax, ``{Same Standards, Different
  Decisions: A Study of QUIC and HTTP/3 Implementation Diversity},'' in
  \emph{Proceedings of the Workshop on the Evolution, Performance, and
  Interoperability of QUIC}, 2020.

\bibitem[Sosnowski et~al.(2023)Sosnowski, Wiedner, Hauser, Steger,
  Schoinianakis, Gallenm{\"u}ller, and Carle]{sosnowski2023performance}
M.~Sosnowski, F.~Wiedner, E.~Hauser, L.~Steger, D.~Schoinianakis,
  S.~Gallenm{\"u}ller, and G.~Carle, ``{The Performance of Post-Quantum TLS
  1.3},'' in \emph{Companion of the 19th International Conference on emerging
  Networking EXperiments and Technologies}, 2023.

\bibitem[Raavi et~al.(2022)Raavi, Wuthier, Chandramouli, Zhou, and
  Chang]{10.1007/978-3-031-22390-7_6}
M.~Raavi, S.~Wuthier, P.~Chandramouli, X.~Zhou, and S.-Y. Chang, ``{QUIC
  Protocol with Post-quantum Authentication},'' in \emph{Information Security},
  2022.

\bibitem[Seemann and Iyengar(2020)]{seemann2020automating}
M.~Seemann and J.~Iyengar, ``{Automating QUIC Interoperability Testing},'' in
  \emph{Proceedings of the Workshop on the Evolution, Performance, and
  Interoperability of QUIC}, 2020.

\bibitem[Zirngibl et~al.(2024)Zirngibl, Gebauer, Sattler, Sosnowski, and
  Carle]{zirngibl2023quic}
J.~Zirngibl, F.~Gebauer, P.~Sattler, M.~Sosnowski, and G.~Carle, ``{QUIC
  Hunter: Finding QUIC Deployments and Identifying Server Libraries Across the
  Internet},'' in \emph{Proc. Passive and Active Measurement (PAM)}, 2024.

\bibitem[König et~al.(2023)König, Waldhorst, and Zitterbart]{koenig2023quic}
M.~König, O.~P. Waldhorst, and M.~Zitterbart, ``{QUIC(k) Enough in the Long
  Run? Sustained Throughput Performance of QUIC Implementations},'' in
  \emph{2023 IEEE 48th Conference on Local Computer Networks (LCN)}, 2023.

\bibitem[{Google}(2024)]{boringssl}
\BIBentryALTinterwordspacing
{Google}, ``{BoringSSL},'' 2024, accessed: 2024-02-29. [Online]. Available:
  \url{https://boringssl.googlesource.com/boringssl}
\BIBentrySTDinterwordspacing

\bibitem[{OpenSSL Project Authors}(2024)]{openssl}
\BIBentryALTinterwordspacing
{OpenSSL Project Authors}, ``{OpenSSL},'' 2024, accessed: 2024-02-29. [Online].
  Available: \url{https://www.openssl.org/}
\BIBentrySTDinterwordspacing

\bibitem[Kempf et~al.(2024)Kempf, Gauder, Jaeger, Zirngibl, and
  Carle]{tls-lib-publication}
\BIBentryALTinterwordspacing
M.~Kempf, N.~Gauder, B.~Jaeger, J.~Zirngibl, and G.~Carle. (2024) {Publication
  of modified TLS Libraries}. [Online]. Available:
  \url{https://github.com/tumi8/quic-crypto-paper}
\BIBentrySTDinterwordspacing

\bibitem[{Open Quantum Safe project}(2024{\natexlab{b}})]{oqs-boringssl}
\BIBentryALTinterwordspacing
{Open Quantum Safe project}, ``{BoringSSL},'' 2024, accessed: 2024-02-29.
  [Online]. Available: \url{https://github.com/open-quantum-safe/boringssl}
\BIBentrySTDinterwordspacing

\bibitem[{National Institute of Standards and Technology}(2024)]{fips203}
\BIBentryALTinterwordspacing
{National Institute of Standards and Technology}, ``{FIPS 203,
  Module-Lattice-Based Key-Encapsulation Mechanism Standard},'' 2024, accessed:
  2024-02-29. [Online]. Available:
  \url{https://doi.org/10.6028/NIST.FIPS.203.ipd}
\BIBentrySTDinterwordspacing

\bibitem[Alagic et~al.(2022)Alagic, Apon, Cooper, Dang, Dang, Kelsey,
  Lichtinger, Liu, Miller, Moody, Peralta, Perlner, Robinson, and
  Smith-Tone]{nist-pqc-status-report-third-round}
\BIBentryALTinterwordspacing
G.~Alagic, D.~Apon, D.~Cooper, Q.~Dang, T.~Dang, J.~Kelsey, J.~Lichtinger,
  Y.-K. Liu, C.~Miller, D.~Moody, R.~Peralta, R.~Perlner, A.~Robinson, and
  D.~Smith-Tone, ``{Status Report on the Third Round of the NIST Post-Quantum
  Cryptography Standardization Process},'' 2022, accessed: 2024-02-29.
  [Online]. Available: \url{https://doi.org/10.6028/NIST.IR.8413-upd1}
\BIBentrySTDinterwordspacing

\bibitem[Stebila et~al.(2023)Stebila, Fluhrer, and
  Gueron]{ietf-tls-hybrid-design-09}
\BIBentryALTinterwordspacing
D.~Stebila, S.~Fluhrer, and S.~Gueron, ``{Hybrid key exchange in TLS 1.3},''
  Internet Engineering Task Force, Internet-Draft
  draft-ietf-tls-hybrid-design-09, Sep. 2023, work in Progress. [Online].
  Available:
  \url{https://datatracker.ietf.org/doc/draft-ietf-tls-hybrid-design/09/}
\BIBentrySTDinterwordspacing

\end{thebibliography}
}

\label{lastpage}

\end{document}